\documentclass[useAMS,usenatbib]{mn2e}
\usepackage{rotating}
\usepackage{journals}

\def\approxgt{\ifmmode \rlap{$>$}{}_{{}_{{}_{\textstyle\sim}}} \else%
$\rlap{$>$}{}_{{}_{{}_{\textstyle\sim}}}$\fi} 
\def\approxlt{\ifmmode \rlap{$<$}{}_{{}_{{}_{\textstyle\sim}}} \else%
$\rlap{$<$}{}_{{}_{{}_{\textstyle\sim}}}$\fi}

\LARGE \normalsize \title[Distances to Galactic LMXBs]{The distances to Galactic low--mass X--ray
  binaries: consequences for black hole luminosities and kicks }

\author[P.G. Jonker \& G. Nelemans]  {P.G. Jonker$^{1,2}$\thanks{email :
    pjonker@head.cfa.harvard.edu}, G. Nelemans$^3$
    \newauthor \\ $^1$Harvard Smithsonian Center for Astrophysics, 60
    Garden Street, MS83, Cambridge, Massachusetts, U.S.A.\\
    $^2$Chandra Fellow\\ $^3$Institute of Astronomy, Madingley Road,
    CB3 0HA, Cambridge, UK\\ }

\begin{document}

\maketitle

\begin{abstract}
\noindent 
We investigated the reported distances of Galactic black hole (BH) and
neutron star low--mass X--ray binaries (LMXBs). Comparing the
distances derived for the neutron stars Cyg~X--2 and XTE~J2123--058
using the observed Eddington limited photospheric radius expansion
bursts with the distances derived using the observed radius and
effective temperature of the companion star we find that the latter
are smaller by approximately a factor of 1.5--2. The latter method is
often employed to determine the distance to BH LMXBs. A possible
explanation for this discrepancy is that the stellar absorption lines
in fast rotating companion stars are different from those in the
slowly rotating template stars as was found before for early type
stars. This could lead to a systematic mis--classification of the
spectral type of the companion star, which in turn would yield a
systematic error in the distance. Further, we derive a distance of
4.0$^{+2.0}_{-1.2}$ kpc for V404 Cyg, using parameters available in
the literature. The interstellar extinction seems to have been
overestimated for XTE~J1550--564 and possibly for two other BH sources
(H~1705--25, and GS~2000+25) as well. As a result hereof the distance
to XTE~J1550--564 may have been underestimated by as much as a factor
three. We find that, using the new distances for XTE~J1550--564 and
V404 Cyg, the maximum outburst luminosity for at least 5 but perhaps
even 7 of the 15 BH soft X--ray transients exceeds the Eddington
luminosity for a $10\,{\rm M_\odot}$ BH showing that these systems
would be classified as ultra--luminous X--ray sources had we observed
them in other Galaxies. This renders support for the idea that many
ultra--luminous X--ray sources are stellar--mass rather than
intermediate--mass BHs. We find that the rms value of the distance to
the Galactic plane for BHs is consistent with that of neutron star
LMXBs. This suggests that BHs could also receive a kick--velocity
during their formation although this has to be investigated in more
detail. We find that the Galactic neutron star and BH l and b
distributions are consistent with being the same. The neutron star and
BH distribution is asymmetric in l with an excess of systems between
-30$^\circ<{\rm l}<0^\circ $ over systems with 0$^\circ<{\rm
l}<30^\circ$.

\end{abstract}


\section{Introduction}
\label{intro}

Low--mass X--ray binaries (LMXBs) are binary systems in which a
compact object -- either a neutron star or a black hole -- accretes
matter from a low--mass companion star. Many of these systems are
found to be transient, so called soft X--ray transients
(SXTs). Several hundreds have been found in our own Galaxy (see
\citealt{2001A&A...368.1021L}) and many of the X--ray point sources in
other Galaxies are likely to be LMXBs as well
(\citealt{1995xrb..book...58V}).  However, some of these X--ray
sources in other Galaxies have luminosities in excess of the Eddington
limits of both neutron star and ten solar mass black hole (BH) LMXBs
(cf.~\citealt{1989ARA&A..27...87F}).  These systems are thought to
contain intermediate mass BHs (\citealt{1999ApJ...519...89C}) or
stellar mass BHs where the emission is anisotropic
(\citealt{2001ApJ...552L.109K}) or indeed super--Eddington
(\citealt{2002ApJ...568L..97B}).

Determining the distance to LMXBs is important, e.g.~to determine the
peak or quiescent luminosity and to determine if the compact object in
LMXBs receives an (asymmetric) kick velocity at birth or
not. Typically, a large distance to the Galactic plane implies the
occurrence of a kick, unless the systems at large z--distances were
formed in the halo.

In this Paper we collect distance estimates from the literature and
discuss possible systematic trends related to the used distance
estimation methods (Section 2). In Section 3 we discuss the
implications for the Galactic distribution of (BH) LMXBs and the
Galactic population of ultra--luminous X--ray sources (ULXs).

\section{Distances to LMXBs}

\subsection{Black hole distances}
\label{sect2.1}
For most BH SXTs it is not feasible to obtain a trigonometric parallax
measurement. Instead the distance generally is determined by comparing
the derived absolute $V$--band magnitude with the (dereddened)
apparent magnitude, taking into account a possible contribution from
residual accretion (the distance derived using this method is
sometimes called a photometric parallax). A first guess of the
distance can be obtained by assuming that the absolute magnitude is
that of a main--sequence star of the observed spectral type, after
determining the best--fit spectral type from the data, e.g.~via the
optimal extraction technique (see \citealt{1994MNRAS.266..137M} for a
description of this technique). We call this method $A$. However,
ideally, the radius, spectral type, and luminosity class are
determined directly from the data. We denote this method $B$. The
observations give the orbital period, P, the radial velocity of the
donor star, K$_2$ and, by comparing the donor star spectrum with
templates that are Doppler broadened, the rotational velocity of the
donor (v\,$\sin i$). The inclination can be estimated from modelling
the ellipsoidal variations in the light curve. From Kepler's third law
and the assumptions that the donor fills its Roche lobe and is in
co--rotation with the orbit, all system parameters can be determined
as follows. The mass function and inclination relate the masses of the
two components, while a combination of Kepler's law and the equation
for the volume of the Roche--lobe gives $\frac{{\rm v\,sin}i}{{\rm
K}_2}=0.46[(1+q)^2q]^\frac{1}{3}$ (e.g.~\citealt{1988ApJ...324..411W}; q
is defined here as the mass of the secondary divided by the mass of
the primary), giving an independent estimate of q thus providing the
system parameters. The radius can be estimated from ${\rm
v\,sin}i=2\,\pi{\rm R\,sin}i/{\rm P}$. Alternatively, if ${\rm
v\,sin}i$ or the inclination are not known, a quite good estimate
of the radius can be obtained using the fact that the mean density of
a Roche--lobe filling star depends only on the orbital period
(\citealt{paczynski1971}). The radius of the donor star can be
estimated via ${\rm R_2 \equiv R_{Roche Lobe}=
0.234\,P_{orb}^{2/3}\,M_2^{1/3}}$ (where ${\rm P_{orb}}$ is in hours,
${\rm M_2}$ in solar masses, and ${\rm R_2}$ in units of solar radii).

In order to estimate the absolute magnitude one either uses the
surface brightness for the observed spectral type or colour, as given
e.g.~by \citet{1976MNRAS.174..489B} and \citet{1980ARA&A..18..115P} or
uses the determined radius and effective temperature, together with an
appropriate bolometric correction. Alternatively, one uses the fact
that for given surface brightness the observed flux scales with the
angular diameter (2 R/d), which together with the radius gives the
distance. However, there are (small) differences in surface brightness
for different luminosity classes of the same spectral type.

In case of GRO~J1655--40 and GRS~1915+105, \citet{1995Natur.375..464H}
and \citet{1999MNRAS.304..865F}, respectively determined limits on the
distance from the observed proper motion for receding and approaching
blobs, assuming the jet ejections are intrinsically symmetric and
noting that the maximum velocity of the ejections is the speed of
light. We list this method as method $C$. The distance can also be
estimated using the interstellar absorption properties of the
source. There are different ways to do this. For several interstellar
absorption lines and diffuse interstellar bands the observed
equivalent width correlates with colour excess
(cf.~\citealt{1995ARA&A..33...19H}). The colour excess can be converted
in a distance estimate (e.g.~using the calibration of
\citealt{1953MNRAS.113..530B}). It is also possible to compare the
observed extinction to that of (OB) stars in the observed field for
which the distance is known or can be derived
(e.g.~\citealt{1986A&AS...63...71V}). The distance can also be
constrained by using high--resolution spectroscopic observations of
the interstellar absorption lines to trace individual gas clouds and
their velocities. Assuming that the apparent velocities of the
different gas clouds are projected Galactic rotation velocities, a
lower limit on the distance of the object can be found
(cf.~\citealt{hynes2004}). We refer to distances derived using the
interstellar absorption properties as method $D$.  Finally, we note
here that various other methods to estimate the distance to BH SXTs
have been proposed and used. We do not discuss these in detail but we
merely mention some of them: \citet{2002MNRAS.331..169H} used the
normalisation of a model describing the accretion disc flux,
\citet{2003A&A...409..697M} used the transition between the high/soft
and low/hard X--ray states, and \citet{2004MNRAS.inpress} proposed to
use the normalisation of the radio -- X--ray correlation in the
low/hard state as a distance indicator.

In Table~\ref{disBHC} we show the distance estimates in the literature
based on method $A$, $B$, $C$, and $D$ for sources for which the best
estimate of the mass of the compact object is above the upper limit of
the neutron star mass of $\sim3$ M$_\odot$ (\citealt{rhoades1974};
\citealt{1996ApJ...470L..61K}). Below, we briefly discuss the sources
listed in Table~\ref{disBHC} for which disparate values for the
distance exist in the recent literature.

\begin{table*}
\caption{BH SXT distance estimates. We indicate whether method $A$,
  $B$, $C$, or $D$ has been used to derive the distance (see
  text). The z--values have been rounded to the nearest 25 pc.}
\label{disBHC}
\begin{center}
\begin{tabular}{lccccccc}
\hline
Name          & l &b  & Spectral & P$_{orb}$ & Distance \& Method & z &  References\\
              & &                & type  & (hours)  & (kpc) & (pc) & \\
\hline
\hline
GRO~J0422+32  & 165.97& -11.99& K9--M2V & 5.09 & 2.5--3.0 ($B$) & -525 - -625& [1,2] \\
1A~0620--00   & 209.96& -6.54& K4--K5V  & 7.75 & 1.2$\pm$0.4 ($B$) &-125 & [3,4] \\
GS~1009--45   & 275.88& 9.35& K6--M0V  & 6.84 & 5.7$\pm$0.7 ($B$) &925 & [5,6] \\
XTE~J1118+480 & 157.66& 62.32& K5--M1V & 4.08 & 1.8$\pm$0.6 ($B$)&1600& [7,8] \\
GS~1124--684  & 295.30& -7.07& K3--K5V & 10.4 & 5.5$\pm$1.0 ($B$)&-675& [9,10] \\
4U~1543--47   & 330.92& 5.43& A2V     & 26.8 & 7.5$\pm$0.5 ($B$)&700 &[11,12] \\
XTE~J1550--564& 325.88& -1.83& G8IV--K4III&37.0& 5.3$\pm$2.3 ($B$)& -175&[13] \\
GRO~J1655--40 & 344.98& 2.46& F2--F6IV & 62.9 & 3.2$\pm0.2$ ($C$)& 125&[14,15,16,17] \\
GX~339--4     & 338.94& -4.33& ??    & 42.1 & $>6$ ($D$)&-450& [18] \\
H~1705--250   & 358.59& 9.06& K0--K5V & 12.5 & 8.6$\pm$2  ($B$)& 1350&[4,19,20] \\
SAX~J1819.3--2525&6.77& -4.79& B9III & 67.6& 9.6$\pm2.4$ ($B$)& -800& [21] \\
XTE~J1859+226 & 54.05& 8.61& G5--K0V & 9.17& $6.3\pm1.7^a$& 950& [22,23] \\
GRS~1915+105  & 45.37& -0.22& K--MIII & 34 (days)& 11$^{+1}_{-4}$ ($C$)& -50& [24,25,26] \\
GS~2000+25    & 63.37& -3.00& K3--K6V & 8.27& 2.7$\pm0.7$ ($B$)& -150&[4,27,28] \\
GS~2023+338   & 73.12& -2.09& G8--K1IV& 6.47 (days) & 4.0$^{+2.0}_{-1.2}$ ($B$)&-150&  [29,30, this work] \\

\end{tabular}
\end{center}

{\footnotesize $^a$ A method to estimate the distance
which has not been discussed here in detail has been used, see
reference 23 for more info. \newline}

{\footnotesize References: [1] \citealt{2003ApJ...599.1254G}, [2]
\citealt{1999A&A...341..491H}, [3] \citealt{1994MNRAS.268..756S}, [4]
\citealt{1996ApJ...473..963B}, [5] \citealt{baolbo2000}, [6]
\citealt{1999PASP..111..969F}, [7] \citealt{2001ApJ...551L.147M}, [8]
\citealt{2001ApJ...556...42W}, [9] \citealt{1996ApJ...468..380O}, [10]
\citealt{esmcna1997}, [11] \citealt{1998ApJ...499..375O}, [12]
\citealt{2002AAS...201.1511O}, [13] \citealt{2002ApJ...568..845O}, [14]
\citealt{1999MNRAS.306...89S}, [15] \citealt{1997ApJ...477..876O}, [16]
\citealt{1995Natur.374..141T}, [17] \citealt{1995Natur.375..464H}, [18]
Hynes et al.~2004, [19] \citealt{1996ApJ...459..226R}, [20]
\citealt{1997AJ....114.1170H}, [21] \citealt{2001ApJ...555..489O}, [22]
\citealt{2001IAUC.7644....2F}, [23] \citealt{2002MNRAS.331..169H}, [24]
\citealt{1994Natur.371...46M}, [25] \citealt{1999MNRAS.304..865F}, [26]
\citealt{2001Natur.414..522G}, [27] \citealt{1996PASP..108..762H}, [28]
\citealt{1996ApJ...470L..57C}, [29] \citealt{1992ApJ...401L..97W}, [30]
\citealt{1994MNRAS.271L..10S}}

\end{table*}

\noindent {\bf 1A~0620--00}:\\ \citet{1994MNRAS.266..137M} report a
distance of 485 pc for 1A~0620--00 whereas most other workers report a
distance close to 1.2 kpc (for references see
Table~\ref{disBHC}). This is due to a numerical error of a factor of 2
in the result of Marsh et al.~(1994; also Marsh private communications).
Hence, the distance Marsh et al.~(1994) find is 912/$\sin i$ pc. The
correct equation for the stellar distance is: 
\begin{equation}
\label{eq1}
d= \frac{2 R}{\theta} =
2\frac{0.46\,q^{1/3}\,(1+q)^{2/3}\,K_2\,P}{2\pi\theta\sin i} = 2
\frac{P (v\sin i)}{2 \pi \theta \sin i}
\end{equation}
Where $\theta$ is the stellar diameter.

\noindent {\bf GRO~J0422+32}:\\ \citet{2000MNRAS.317..528W} report a
distance of 1.39$\pm$0.15 kpc. However, besides the fact the best--fit
spectral type they derived for the companion star was M4--5, later
than that derived by e.g.~\citet{1999A&A...341..491H} and
\citet{2003ApJ...599.1254G} who found a M1--2 spectral
type, they used method $A$ which is known to underestimate the radius
of the companion star. In the distance determination
\citet{2000MNRAS.317..528W} use the absolute magnitude of a star with
radius R = 0.24 R$_\odot$ (from
\citealt{1993ApJ...402..643K} via the bolometric magnitude and the
effective temperature), whereas the radius one would derive using the
equation in the first paragraph of Section~\ref{sect2.1} would give R
= 0.53 R$_\odot$.

For that reason, we prefer the distance derived using method $B$ by
\citet{1999A&A...341..491H} and \citet{2003ApJ...599.1254G}.  
However, \citet{1999A&A...341..491H} used the same erroneous equation
as Marsh et al.~(1994). If we correct for this error,
\citet{1999A&A...341..491H} found a distance of 5.0$\pm$1.6 kpc.
Interestingly, \citet{2003ApJ...599.1254G} found a distance of
2.5$\pm$0.3 kpc even though they also used method $B$. This is due to
the fact that \citet{1999A&A...341..491H} use $(V-R)_0\sim1$ for an
M1--2 star, while, both in
\citet{1984AcA....34...97E} and \citet{2000asqu.book.....C} give a value for 
the $(V-R)_0$ colour closer to 1.5 for a M2V star. Taking a $(V-R)_0$
of 1.5 for an M2 as in the relation of \citet{1984AcA....34...97E}
would yield a distance close to 2.5 kpc. Indeed, if we take the radius
of the secondary and apparent $V$--band magnitude corrected for
interstellar extinction and an accretion disc contribution as given by
\citet{1999A&A...341..491H} but use the relation given by
\citet{1980ARA&A..18..115P} for an M2 star we find a distance of 2.8
kpc. We conclude that the distance to GRO~J0422+32 is likely to be
2.5--3.0 kpc.

\noindent {\bf GX~339--4}:\\ In the case of GX~339--4, the distance
has been estimated to be $\sim4$ kpc using the equivalent width of the
CaII--K interstellar line (e.g.~\citealt{1987AJ.....93..195C}; see
\citealt{2003MNRAS.342..105B} for an overview). Recently, from high
resolution spectroscopic observations resolving the contributions to
the interstellar Na~D absorption lines, \citet{hynes2004} found that
the distance to GX~339--4 is likely to be more than 6 kpc. They explain
that in order for the distance to be $\sim4$ kpc the line--of--sight
towards GX~339--4 must be peculiar.  The limit on the distance of
GX~339--4 found by \citet{2003A&A...409..697M} indicates that 
d$>$7.6 kpc. In Table~\ref{disBHC} we refer to the value derived by
\citet{hynes2004}.

\noindent {\bf GS~2023+338, a.k.a.~V404~Cyg}:\\ The distance to
GS~2023+338 has been reported to be close to 3 kpc
(\citealt{1994MNRAS.271L..10S}) or 8 kpc (\citealt{1996ApJ...473L..25W};
even a distance of 11 kpc has been mentioned
\citealt{1992Natur.355..614C}). \citet{1993MNRAS.260L...5K} finds that
if the secondary is a stripped--giant, GS~2023+338 must have a
distance 3.5--5.1 kpc.  In an attempt to reconcile the different
distances we recalculate the distance using method $B$. We used the
relation between the absolute visual magnitude, spectral type and
radius of the companion star given by \citet{1980ARA&A..18..115P}. The
spectral type of GS~2023+338 is K0$\pm$1
(\citealt{1993MNRAS.265..834C}). We take ${\rm R_2=6\pm1 R_\odot}$ for
the radius of the companion star after \citet{1994MNRAS.271L..10S} who
obtained this from modelling the ellipsoidal variations in the
$K$--band light curve. From the relation of
\citet{1980ARA&A..18..115P} we find for the absolute magnitude
$1.8\leq {\rm M_V} \leq 3.0$.  We further take $m_v=18.7$ using the
observed $V$--band magnitude of 18.42 and the fact that approximately
25 per cent of the light in the $V$--band was estimated to come from
the accretion disc, although the uncertainty in the accretion disc
contribution is large (\citealt{1993MNRAS.265..834C}). Here we took
25$\pm$15 per cent for the accretion disc contribution to the light in
the $V$--band; from this we get for the observed $V$--band magnitude
$18.5\leq {\rm m_v} \leq 19.0$. Together with an assumed interstellar
absorption of ${\rm A_V}=3.3$, this yields a distance of
4.0$^{+2.0}_{-1.2}$ kpc. Note that this calculation does not include
an uncertainty in the interstellar absorption (see below). Since, as
mentioned above, the accretion disk contribution is uncertain we also
determined the distance assuming that there is no contribution of the
accretion disc to the $K$--band flux (this assumption yields a lower
limit to the distance). For a K0V/III star the $(V-K)_0$ colour is
1.96/2.31 (\citealt{2000asqu.book.....C}), using the relations between
the $(V-K)_0$ colour and the surface brightness as determined by
\citet{1981MNRAS.197...31B} we get a lower limit to the distance 
of 2.7 kpc for a K0III and 3 kpc for a K0V companion star (again we
used ${\rm R_2=6\pm1 R_\odot}$ after \citealt{1994MNRAS.271L..10S}) and
we took A$_K=0.4$.

\subsection{Neutron star LMXB distances}

In order to compare the BH distances with the neutron star LMXB
distances, we list in Table~\ref{NSdist} the distances to neutron star
LMXBs. We excluded sources in Globular Clusters since we want to
compare the neutron star sample with that of BHs and (so far) BHs have
not been found in Globular Clusters.

Some of the type~I X--ray bursts, more specifically those showing
evidence for photospheric radius expansion can be used as a standard
candle (\citealt{1978Natur.274..650V}). Using the distance of the
Galactic Centre and those of Globular Clusters to estimate the maximum
peak luminosity for photospheric radius expansion bursts
\citet{1981A&A...101..174V} found that the average peak luminosity of
photospheric radius expansion bursts is $3\times10^{38}$ erg s$^{-1}$.
\citet{1984MNRAS.210..899V} used a more extensive sample of X--ray
bursters in Globular Clusters and found an average peak luminosity of
$(4.0\pm0.9)\times10^{38}$ erg s$^{-1}$. Recently,
\citet{2003A&A...399..663K} found a neutron star Eddington luminosity
of $(3.79\pm0.15)\times10^{38}$ erg s$^{-1}$ for the peak luminosity
of radius expansion bursts in Globular Clusters. This luminosity is
consistent with the Eddington luminosity of a 1.4 M$_\odot$ neutron
star accreting helium rich material. \citet{2003A&A...399..663K} also
found that a few systems have a lower peak luminosity of
$\sim2\times10^{38}$ erg s$^{-1}$ which can be interpreted as the
Eddington luminosity for hydrogen rich accreted material.  Therefore,
and for reasons explained in Section \ref{nsuncer}, we use both a peak
luminosities to calculate the distance to the neutron star LMXB.

For Aql~X--1 we determined the burst peak flux in a 0.25~s bin from a
photospheric radius expansion burst detected with the {\it RXTE}
satellite in the observation starting on MJD 51364.834069 (Terrestrial
Time). In this we subtracted the average persistent emission 5--105
seconds before the burst. The bolometric burst peak flux is
1.1$\times10^{-7}$ erg cm$^{-2}$ s$^{-1}$. We corrected the Aql~X--1
flux for the fact that fluxes derived using the RXTE satellite are
found to be systematically higher by about 20 per cent than the X--ray
fluxes in the same band found by other satellites
(cf.~\citealt{1999ApJ...512..892T}; \citealt{baolbo2000};
\citealt{2003A&A...399..663K}). Finally, we use the distance for
Sco~X--1 as determined from radio parallax measurements (d=2.8$\pm$0.3
kpc, \citealt{1999ApJ...512L.121B}).

\begin{table*}
\caption{Properties of the sample of neutron stars used in this
Paper. Distances derived from type~I photospheric radius expansion
bursts in both persistent and transient neutron star systems
(excluding those in Globular Clusters) using a neutron star Eddington
luminosity of 2.0 or 3.8$\times10^{38}$ erg s$^{-1}$. The burst peak
flux (unabsorbed, 0.1--100 keV; indicated with a {\it i} behind the
flux) or the bolometric peak burst flux (indicated with a {\it ii}
behind the flux) is given as well. The z--values are rounded to the
nearest 25 pc.}
\label{NSdist}
\begin{center}

\begin{tabular}{lcccccccc}
\hline
Name  & l  & b  & T/P$^a$ & P$_{orb}$& Peak burst flux & Distance & z & References\\
 & & & & hours &  erg cm$^{-2}$ s$^{-1}$ & 2.0--3.8$^c$ (kpc) & 2.0--3.8$^c$ (pc) & \\
\hline
\hline
EXO~0748--676& 279.98 & -19.81 & T & 3.82 & 3.8$\times10^{-8}$ ({\it i})& 6.8--9.1&-2300 -- -3075 & [1] \\
2S~0918--54 & 275.85 &-3.84&  P &  ?  & 9.4$\times10^{-8}$ ({\it ii})& 4.3--5.8&-300 -- -400& [18,19] \\
Cir~X--1     & 322.12 & 0.04&  T/P?& 398 & 2.9$\times10^{-8}$ ({\it i})& 7.8--10.5&0 -- 0&  [2,3] \\
4U~1608--522 & 330.93 & -0.85&  T  & 12.9? & 2.2$\times10^{-7}$ ({\it i})& 2.8--3.8&-50 -- -50 & [4] \\
Sco~X--1    & 350.09 & 23.78 & P & 18.9  & parallax & 2.8$\pm0.3$   & 1125 & [20]\\
4U~1636--53 & 332.91 &-4.82 &  P &  3.80 & 1.3$\times10^{-7}$ ({\it i})& 3.7--4.9&-300 -- -400& [21] \\
4U~1658--298 & 353.83 & 7.27& T  & 7.11 & 2.5$\times10^{-8}$ ({\it ii})& 8.4--11.3&1075 -- 1425 & [5] \\
4U~1702--429 & 343.89 & -1.32 & P & ? & 6.6$\times10^{-8,b}$ ({\it ii}) & 5.3--7.1 &-125 -- -175 & [32]\\
4U~1705--44 & 343.32 &  -2.34 & P & ? & 3.7$\times10^{-8,b}$ ({\it ii}) & 7.2--9.6 &-300 -- -400 & [32]\\
XTE~J1710--281 & 356.36 & 6.92& T & ? & 8.6$\times10^{-9,b}$ ({\it ii}) & 14.8--19.8 &1800 -- 2375 & [32]\\
SAX~J1712.6--3739 &348.93&0.93 & T& ? & 5.1$\times10^{-8}$ ({\it ii})& 5.9--7.9&100 -- 125 & [6] \\
1H~1715--321& 354.13 &3.06 &  P/T?& ? & 6.7$\times10^{-8}$ ({\it ii})& 5.1--6.9&275 -- 375& [22] \\
RX~J1718.4--4029&347.28&-1.65 & P/T?& ? &4.3$\times10^{-8}$ ({\it i})& 6.4--8.6&-200 -- -250& [23] \\
4U~1728--34&354.30&-0.15 &  P &  ?  &  8.4$\times10^{-8}$ ({\it i})& 4.5--6.1&0 -- -25& [24,25] \\
KS~1731--260 & 1.07 & 3.65& T  & ? & 6.3$\times10^{-8}$ ({\it ii})& 5.3--7.1&325 -- 450 & [7] \\
4U~1735--44 & 346.05&  -6.99 & P & ? & 2.9$\times10^{-8,b}$ ({\it ii}) & 8.0--10.8 &-975 -- -1325 & [32]\\
GRS~1741.9--2853 & 359.96& 0.13& T& ? &  4.0$\times10^{-8}$ ({\it i})& 6.6--8.9&25 -- 25 & [8] \\
2E~1742.9--2929 & 359.56 & -0.39& T/P? & ? &3.7$\times10^{-8,b}$ ({\it ii}) & 6.9--9.2 &-50 -- -75 & [32]\\
SAX~J1747.0--2853& 0.21 & -0.24& T& ? & 3.2$\times10^{-8}$ ({\it ii})& 7.5--10&-25 -- -50 & [9] \\
GX~3+1     &2.29&0.79&  P &  ?    &  9.3$\times10^{-8}$ ({\it i})& 4.3--5.8&50 -- 75& [26] \\
SAX~J1750.8--2900& 0.45 & -0.95& T& ? &  6.4$\times10^{-8}$ ({\it ii})& 5.2--7.0&-75 -- -125& [10] \\
SAX~J1752.3--3138&358.44&-2.64 & P/T?& ? & 2.8$\times10^{-8}$({\it ii})&7.9--10.6&-375 -- -475& [27] \\
SAX~J1808.4--3658& 355.38& -8.15 & T& 2.0&  2.5$\times10^{-7}$ ({\it ii})& 2.7--3.6&-375 -- -500 & [11] \\
SAX~J1810.8--2609& 5.20 &-3.43& T& ? & 7.0$\times10^{-8}$ ({\it ii})& 5.1--6.8&-300 -- -400& [12] \\
4U~1812--12&18.06 &2.38&  P &  ?    &  1.5$\times10^{-7}$ ({\it ii})& 3.4--4.6&150 -- 200& [28] \\
XTE~J1814--338 & 358.75 &-7.59 &T& 4.27  & 2.6$\times10^{-8}$ ({\it ii})& 8.2--11.0&-1075 -- -1450  & [13] \\
GX~17+2&16.43&1.28&  P     & ? &  1.2$\times10^{-8}$ ({\it ii})& 11.9--16.0&275 -- 350& [29] \\
Ser~X--1 & 36.12 & 4.84& P & ? & 2.1$\times10^{-8,b}$ ({\it ii}) & 9.5--12.7 &800 -- 1075 & [32]\\
Aql~X--1 & 35.72 &-4.14 & T    & 19.0  & 9.1$\times10^{-8,b}$ ({\it ii})& 4.4--5.9&-325 -- -425& [14] \\
4U~1857+01 & 35.02 &-3.71 &  T  & ?  &   3$\times10^{-8}$ ({\it ii})& 7.5--10&-500 -- -650& [15] \\
4U~1916--053&31.36&-8.46&  P & 0.83  & 3.1$\times10^{-8}$ ({\it ii})& 7.5--10.1&-1100 -- -1475& [30] \\
XTE~J2123--058 &46.48&-36.20 & T & 5.96 & 7$\times10^{-9}$ ({\it ii})& 15.7--21&-9275 -- -12.4$\times10^3$& [16,17] \\
Cyg~X--2&87.33&-11.32&  P & 236.2    &     1.35$\times10^{-8}$({\it ii})& 11.4--15.3&-2250 -- -3000& [31] \\
\end{tabular}
\end{center}
{\footnotesize $^a$ T= transient, P=persistent } \newline
{\footnotesize $^b$ Corrected for the fact that RXTE fluxes are found to be higher by 20 per cent } \newline
{\footnotesize $^c$ Determined assuming an Eddington peak flux of 2.0 or 3.8$\times 10^{38}$ erg s$^{-1}$ } \newline

{\footnotesize References: [1] \citealt{gohapa1986}, [2]
\citealt{1986MNRAS.221P..27T}, [3] \citealt{1996MNRAS.283.1071B}, [4]
\citealt{1987PASJ...39..879M}, [5] \citealt{2002ApJ...566.1060W}, [6]
\citealt{2001MmSAI..72..757C}, [7] \citealt{2001ApJ...553L.157M}, [8]
\citealt{1999A&A...346L..45C}, [9] \citealt{2000ApJ...543L..73N}, [10]
\citealt{2002ApJ...575.1018K}, [11] \citealt{2001A&A...372..916I}, [12]
\citealt{2000ApJ...536..891N}, [13] \citealt{2003ApJ...596L..67S}, [14]
This work, [15] \citealt{1990A&A...228..115C}, [16]
\citealt{1999ApJ...513L.119H}, [17] \citealt{1999ApJ...521..341T}, [18]
\citealt{2001ApJ...553..335J}, [19] \citealt{2002A&A...392..885C}, [20]
\citealt{1999ApJ...512L.121B}, [21] \citealt{1988A&A...199L...9F} and references therein, [22]
\citealt{1984ApJ...276L..41T}, [23] \citealt{2000A&A...358L..71K}, [24]
\citealt{1984ApJ...281..337B}, their ``super--burst'', [25]
\citealt{1980ApJ...240L..27H}, [26] \citealt{2000A&A...356L..45K}, [27]
\citealt{2001A&A...378L..37C}, [28] \citealt{2000A&A...357..527C}, [29]
\citealt{2002A&A...382..947K}, [30] \citealt{2001ApJ...549L..85G}, [31]
\citealt{1998ApJ...498L.141S}, [32] Galloway et al.~2004, in prep}\newline

\end{table*}

\subsection{Systematics and uncertainties in distance estimates}

\citet{1999MNRAS.305..132O} derived a distance of 7.2$\pm1.1$ kpc for
Cyg~X--2, using method $B$ (they used the spectral type of the
companion derived by \citealt{1998ApJ...493L..39C}). From the radius
expansion burst (\citealt{1998ApJ...498L.141S}) we find a distance of
15.3 kpc, approximately a factor two larger.  The large differences in
these distance estimates are difficult to explain. Even if the radius
expansion bursts are hydrogen--rich and hence the burst peak
luminosity is lower (see Section~\ref{nsuncer}), there would still be
a difference in distance estimates of a factor 1.5. However, due to
the large photon rate deadtime effects could lower the apparent burst
peak flux. The bolometric burst peak flux corrected for deadtime
effects was 1.5$\times10^{-8}$ erg cm$^{-2}$ s$^{-1}$
(\citealt{1998ApJ...498L.141S}; compare this with the value in
Table~\ref{NSdist}). This makes the distance smaller by $\sim$4 per
cent. On the other hand, correcting the observed fluxes for the fact
that the RXTE/PCA fluxes are generally found to be 20 per cent higher
than fluxes determined using other satellites would make the radius
expansion burst distance larger. Furthermore, if Cyg~X--2 is a halo
object with a low metallicity the absolute $V$--band magnitude would
be smaller than that of a star with solar metallicity in order to
explain the observed spectral type, making the discrepancy even
bigger. A possible solution for the distance discrepancy could be that
the neutron star in Cyg~X--2 is lighter than the canonical value of
1.4 M$_\odot$ (although this would also affect the optically
determined distance). However, this is at odds with the findings of
\citet{1998ApJ...493L..39C} and \citet{1999MNRAS.305..132O} who find
that the mass of the neutron star in Cyg~X--2 is $>$1.88 M$_\odot$ and
1.78$\pm$0.23 M$_\odot$, respectively.

From spectroscopic observations of XTE~J2123--058 in quiescence
\citet{2002MNRAS.329...29C} determined a best--fit spectral type for
the companion star of K7V (they ruled out spectral types earlier than
K3 and later than M1). \citet{2003ApJ...585..443S} report a quiescent
$V$--band magnitude for XTE~J2123--058 of 22.65$\pm$0.06, a reddening
in the $V$--band of 0.37 magnitudes, and that the companion star
contributes approximately 77 per cent of the flux in the
$R$--band. Here we assume that the companion star contributes 70 per
cent of the flux in the $V$--band. Again using the relation of
\citet{1980ARA&A..18..115P} to estimate the absolute $V$--band
magnitude gives ${\rm M_V=7.97}$ for the observed mass ratio of 0.49
and assuming a neutron star mass of 1.4 M$_\odot$ (had we assumed a
neutron star mass of 2.0 M$_\odot$ we would have gotten ${\rm
M_V=7.72}$). We again assumed that the companion star fills its Roche
lobe. From this we derive a distance of 8.7/9.6 kpc for a neutron star
mass of 1.4/2.0 M$_\odot$, respectively (see also
\citealt{2004ApJTomsick2123} and references therein). The distance derived 
using the radius expansion burst is 15.7--21 kpc (see
Table~\ref{NSdist}; deadtime effects are negligible in case of
XTE~J2123--058). For the limiting spectral types of the companion star
(K3/M0V) the distance would be 14/7 kpc for a 1.4 M$_\odot$ neutron
star. So, even for an assumed K3 spectral type the distance is lower
than that derived using the radius expansion burst. Again, as in the
case of Cyg~X--2, correcting the parameters we used in the distance
calculations either for the fact that the RXTE/PCA fluxes are
generally found to be 20 per cent higher than fluxes determined using
other satellites or for the fact that XTE~J2123--058 might be a halo
object with a low metallicity would make the discrepancy between the
photospheric radius expansion burst--distance and the method $B$
bigger.

Even though the sample of neutron star sources for which we can
compare the distances derived using radius expansion bursts and that
derived using the properties of the companion star is small (so far
this is only possible for two source), this could mean that distances
derived using method $B$ are too low, or that the distance derived
using the radius expansion burst is too large. Below we will
investigate in some detail possible effects responsible for the
observed discrepancy in distance estimate between method $B$ and the
radius expansion burst method.

\subsubsection{Systematics in the companion star radius determination and
 residual accretion}

As mentioned earlier, under the assumption that the companion star
fills its Roche lobe a first estimate of the radius of the companion
star can be obtained (cf.~\citealt{1996ApJ...473L..25W}). Given the fact
that mass accretion must take place in order to explain the multiple
outbursts for some of the systems the assumption that the secondary
(nearly) fills its Roche lobe seems justified. However, if the
temperature distribution on the surface of the star is uneven e.g.~due
to effects of irradiation, then the luminosity is not determined by
the full size of the companion star. Depending on whether the
determined temperature is that of the hotter or colder part of the
star this yields an over or under estimate of the distance. 

An under estimate of the amount of light contributed due to residual
accretion would lead to an under estimate of the distance and vice
versa.

\subsubsection{Systematics in the companion star temperature determination}

\begin{figure}
\includegraphics[width=7cm,angle=-90]{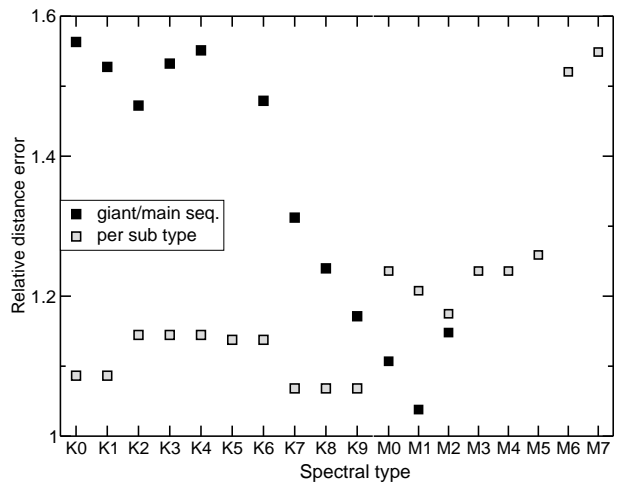}
\caption{The fractional error one would make in the distance estimate
if the spectral classification is off by 1 sub--class (grey squares)
is plotted as a function of spectral type for late type stars. The
fractional error in the distance estimate if one takes a
main--sequence spectral class companion instead of a giant of the same
spectral type for the same radius (black, filled squares). For both it
is assumed that the radius of the companion star is known and accurate
(in the calculations we used the results from Popper 1980).}
\label{fracerr}
\end{figure}

Besides the radius, the temperature of the companion star is important
for its luminosity. The temperature is derived from the observed
spectral type.  A systematic mis--classification of one spectral type
for a fixed radius (e.g.~using a K1V instead of a K0V star) already
results in a distance error of $\sim$15--25 per cent for late type
stars (see Figure~\ref{fracerr}). The difference in derived distance
between using the surface brightness of main--sequence stars or giants
(for fixed radius) is also given in Figure~\ref{fracerr}. The giant
surface brightnesses lead to smaller absolute magnitudes, i.e.~smaller
distances.

The best--fit spectral type can be determined using the optimal
subtraction method (\citealt{1994MNRAS.266..137M}) which is quite
robust. However, this method is rarely applied fully.  The broadband
spectral energy distribution can also be used to determine the
spectral type (cf.~\citealt{2001AJ....122..971G} and
\citealt{2003ApJ...599.1254G} for recent use of this method). However,
disentangling the effects of fast rotation, reddening, a possible
accretion disc contribution, and the spectral type is difficult with
only a few data points. This makes the uncertainty in the temperature
of the companion star an important contributor to the uncertainty in
the distance. For XTE~J2123--058 a spectral type of K2 (just outside
the determined range) would already make the two different distance
estimates consistent. For Cyg~X--2 a spectral type of A0 is needed
while the determined type is A9$\pm$2. However, from the effects
discussed above, there is no reason to expect a systematic
underestimation of the effective temperature of the companion star
which would lead to a systematic underestimation of the distance using
method $B$.

In using method $B$ as described above an implicit assumption about
the temperature of the companion star has been made. This is due to
the fact that the effective temperature of the companion star is in
most cases assumed to be that of a main--sequence star of the observed
spectral type. Besides a small variation in temperature with
luminosity class there is also a difference in bolometric correction
for stars which have the same spectral type but a different luminosity
class. Comparing the observed surface gravity with that of different
luminosity classes of the observed spectral type the effective
temperature of the companion star can be determined. These then
determine the bolometric correction, and thus the distance. However,
the error in the distance from neglecting the difference in bolometric
correction between the luminosity classes is small.

However, it is possible that the observed spectral types are
systematically shifted to later spectral types because of effects on
the lines and continuum induced by the fast rotation of the companion
star. Single late type stars have rotational velocities of less than a
few kilometres per second due to the onset of a magnetic--brake in
stars later than typically F5 (see
\citealt{1967ApJ...150..551K}). However, the late type companion stars
in LMXBs are likely in co-rotation with the orbit. Therefore, the
rotational velocities range from several tens of km s$^{-1}$ for long
orbital period systems (e.g.~GRS~1915+105;
\citealt{2004A&A...414L..13H}) to hundred or more km s$^{-1}$ for the
short orbital period systems. Studies of rapidly rotating early type
stars found that a fast rotation causes a slight increase in absolute
magnitude of stars with spectral type later than B5 (the stars are
intrinsically less luminous than the non--rotating stars of the same
spectral type by typically several tenths of a magnitude; only stars
up F8 were studied; \citealt{1977ApJS...34...41C}). Additionally, it is
known that limb and gravity darkening effects on the line change the
equivalent widths of stellar photospheric absorption lines
(e.g.~\citealt{1929MNRAS..89..222S};
\citealt{1977ApJS...34...41C}).\footnote{e.g.~the ratio between
equivalent widths of the hydrogen lines and those of weak metallic
atoms also changes among other things due to the fact that the strong
hydrogen lines respond different to the rotational effects
(\citealt{1991ApJS...77..541C}). \citet{2001ApJ...555..489O} found that
the companion star in V4641~Sgr (a.k.a.~SAX~J1819.3--2525) is rapidly
rotating (v~sin$i\sim$123 km s$^{-1}$). In fact, for an assumed B9III
spectral type observed at an inclination close to 60$^\circ$
(\citealt{2001ApJ...555..489O}), the star is rotating at 80--90 per cent
of its break--up speed (cf.~\citealt{1977ApJS...34...41C}). The fact
that the lines of stars with large rotational velocities are similar
to the lines in stars of later spectral types
(\citealt{1977ApJS...34...41C}) can explain the unusually strong Mg~II
line in V4641~Sgr. \citet{2000ApJ...541..908B} noted that the peculiar
Li abundances observed in the companion stars of several BH SXTs could
also result from the fact that the companion stars are fast rotators
similarly to RS~CVn stars.}  Generally, the lines in the spectrum
resemble the lines of a later spectral type star
(\citealt{1977ApJS...34...41C}). However, this depends on the behaviour
of the equivalent width of the used lines as a function of the
effective temperature.  Broadening of template star spectra in the
optimal subtraction technique does not take into account these
physical changes in the lines. In early type stars the decrease in
luminosity of the star due to rotation is outweighed by the apparent
shift in spectral type. Hence, the distance estimation method $B$
using spectra of non--rotating template stars to estimate the spectral
type applied to fast rotating early type stars would likely lead to a
net underestimation of the distance, the effect being bigger the later
the spectral type of the star (at least up to spectral type F8).

Unfortunately, there are no detailed model atmosphere calculations
showing the physical effects of rotation on the absolute magnitude and
the lines of stars later than F8 available in the literature. However,
effects of limb darkening are larger in late type stars than in early
type stars. Furthermore, extrapolating the findings for the early type
stars the spectral appearance is likely more affected than the true
source luminosity meaning that the distances to the sources would be
underestimated using non--rotating template stars to determine the
spectral type.

\subsubsection{Systematics in the interstellar absorption estimates}
If the interstellar extinction, ${\rm A_V}$, is systematically
overestimated this would yield an underestimate of the distance or
vice versa since ${\rm m_V - A_V - M_V = 5\,log}\,d - 5$. The
interstellar extinction caused by scattering of light off dust grains
is traced by the optical/UV spectrum (through E(B--V) and/or the
strength of the absorption feature at 2175\AA).  Spectral fits to the
X--ray data trace interstellar absorption caused by neutral hydrogen
and absorption (above $\sim$0.5 keV) by K/L--shell electrons of mostly
O and Fe (\citealt{1995A&A...293..889P}). Finally, interstellar
absorption lines from ions such as Na~I and Ca~II trace
low--ionisation interstellar gas (\citealt{1994A&A...289..539S}).  Over
the years relations between ${\rm A_V}$ and these different sources of
interstellar extinction have been
established. E.g.~\citet{1995A&A...293..889P} reported a relation
between ${\rm N_H}$ and ${\rm A_V}$. \citet{1985ApJ...288..618R} give
a relation between ${\rm E(B-V)}$ and ${\rm A_V}$ (${\rm R
\times E(B-V) = A_V}$, with R$\approx$3.1).  \citet{1995ARA&A..33...19H} 
gives relations between the equivalent widths of the Na~I and Ca~II
absorption lines and ${\rm E(B-V)}$. However, the line--depth of some
absorption lines saturates quickly leading to an overestimate of the
distance. Furthermore, for systems out of the Galactic plane the
equivalent width of the interstellar absorption features does not
increase much with distance once the system has a z--distance larger
than the scale--height of the interstellar material (e.g.~the
scale--height of Ca~II is approximately 800~pc whereas that of Na~I is
approximately 400--500 pc; \citealt{1994A&A...289..539S}).

In addition, determining ${\rm A_V}$ from the observed ${\rm N_H}$ or
from the observed equivalent width of interstellar absorption lines
one implicitly assumes that the sight--line for the SXT under
investigation has the same gas--to--dust ratio (i.e.~${\rm N_H}$/${\rm
E(B-V)}$) as the sight--lines for which the relations between ${\rm
A_V}$ and ${\rm N_H}$ or between the equivalent widths of the Na~I and
Ca~II absorption lines and ${\rm E(B-V)}$ have been established. It is
known that this is not always the case; e.g.~the sight--line towards
the Gould belt (i.e.~Orion, Ophiuchus, and Perseus;
\citealt{1978ApJ...225...40B}) has a low gas--to--dust ratio. Similarly,
the sight--line towards the Galactic Centre, where most LMXBs are
located, has, in general, a low R (R$<$3.1;
\citealt{2003ApJ...590..284U}). This caveat has to be kept in mind.

In Table~\ref{av} we show the ${\rm A_V}$ as derived from optical/UV
and X--ray data.  As was found before
(e.g.~\citealt{1991ApJS...76.1127V}), the extinction derived using
X--ray data is systematically larger than that based on optical/UV
data. There are two possible reasons for this.
\citet{1995A&A...293..889P} used the ${\rm N_H}$ values derived from
fitting an absorbed power--law model to ROSAT X--ray data. Using an
absorbed Bremsstrahlung or a black body model to fit the data gave
systematically lower values for ${\rm N_H}$. However, in the ${\rm
N_H}$ estimates in Table~\ref{av} in most cases a power--law model has
been used as well. 

Another possibility is that local absorbing material is present during
outbursts. Since most of the X--ray observations which are sensitive
enough to determine ${\rm N_H}$ are done during outburst this could
lead to a systematic overestimation of ${\rm N_H}$. The fact that for
XTE~J1859+226 the value derived from X--ray spectral fits is higher
than that derived using the UV data even though both the HST UV and
the BeppoSAX X--ray observations were made during outburst can be
explained by the fact that the UV and X--ray observations are
sensitive to different sources of interstellar extinction. If the
extra absorbing material is indeed local to the source, the
temperature may be too high for dust to form, on the other hand
neutral hydrogen may well be present explaining the difference in
derived ${\rm A_V}$.  This all suggests that the ${\rm A_V}$ derived
from ${\rm N_H}$ from X--ray spectral fitting, systematically provides
an overestimate for ${\rm A_V}$. Hence, their use would yield an
underestimation of the distance.

In all BH distance estimates in the literature the ${\rm A_V}$ derived
using optical or UV data has been used, except in case of
XTE~J1550--564, H~1705--25, and GS~2000+25. The diffuse interstellar
bands (DIBs) used by \citet{1999A&A...348L...9S} to estimate the ${\rm
A_V}$ for XTE~J1550--564 do not suffer from saturation at relatively
low values of ${\rm E(B-V)}$. Given the fact that the factor with
which the distance to the source changes goes as
$\frac{d_2}{d_1}=10^{\frac{A_{V1}-A_{V2}}{5}}$, the distance to
XTE~J1550--564 may have been underestimated by a factor $\sim$3!
Although, as explained above, the sight--line could have properties
different from the properties of the sight--line used in deriving the
relation between the ${\rm A_V}$ and the equivalent width of the DIB
leading to an anomalous low ${\rm A_V}$ as derived from the DIB. In
case of GS~2000+25 an ${\rm A_V}$ of 4.4 has been used. However, since
the optically derived value for ${\rm A_V}$ for GS~2000+25 does not
seem to be very accurate (it has been determined assuming the
intrinsic B-V colour of the source during outburst is 0) it is unclear
whether this is an overestimation or not. Unfortunately, it is not
possible to estimate whether the distance to H~1705--225 is over or
underestimated since there is no optically derived ${\rm A_V}$
available.

\begin{table}
\caption{The interstellar extinction ${\rm A_V}$ for the sample of BH
SXTs determined using optical/UV or X--ray data. For some sources a
range of values corresponding to the different values found by the
different references is given whereas for others an error bar is
given. We used ${\rm R=3.1}$ in ${\rm A_V=R \times E(B-V)}$. We used
the $N_H$ value obtained using a power--law model where available
since the correlation ${\rm N_H =1.79\times10^{21}\,cm^{-2}\,A_V }$
was derived using power--law fits to ROSAT data (see Predehl \&
Schmidt 1995). An ``X'' denotes that the value has not been determined. }
\label{av}

\begin{center}
\begin{tabular}{lccc}

\hline
Name          & ${\rm A_V}$ optical   & ${\rm A_V}$ X--ray & 
Refs\\
 & & & Opt -- X--ray\\
\hline
\hline
GRO~J0422+32  & 0.6--1.2$^a$  & $<$2.8 & [1] -- [2]\\
1A~0620--00   & 1.09--1.24  & 1.3$\pm$0.7 & [3,4] -- [5]\\
GS~1009--45   & 0.6$\pm$0.2 & X    & [6] -- [X]\\
XTE~J1118+480 & X     & 0.041$\pm$0.004$^d$   & [X] -- [7]\\
GS~1124--684  & 0.9$\pm$0.1  & 1.28$\pm$0.06 & [8] -- [9]\\
4U~1543--47   & 1.55$\pm$0.15& 2.4$\pm$0.1 & [10] -- [11]\\
XTE~J1550--564& 2.5$^e$& 4.88$\pm$1.15 & [12] -- [13]\\
GRO~J1655--40 & 3.7$\pm$0.3  & 4.8$\pm$2.8 & [14] -- [5]\\
GX~339--4     & $>$2.8 & 3.9$\pm$0.5 & [15] -- [16]\\
H~1705--250   & X     & 1.7$\pm$0.5  & [X] -- [17]\\
SAX~J1819.3--2525&1.0$\pm$0.3$^b$& X& [18] -- [X]\\
XTE~J1859+226 & 1.80$\pm$0.37& 4.47$^c$ & [19] -- [20]\\
GS~2000+25    & 3.5$^b$  & 6.4$\pm$1.0 & [21] -- [22]\\
GS~2023+338   & 3.3--4.0 & 3.9$\pm$0.4 & [23,24] -- [5]\\
\hline
\end{tabular}
\end{center}
{\footnotesize $^a$ see the discussion and references in \citet{1997MNRAS.290..303B}} \newline 
{\footnotesize $^b$ the uncertainty is large since the value is derived assuming
$(B-V)_0=0$ } \newline 
{\footnotesize $^c$ no error bars given } \newline
{\footnotesize $^d$ ${\rm N_H}$ from EUVE observations } \newline
{\footnotesize $^e$ \citet{1999A&A...348L...9S} report 2.2$\pm$0.3. We
prefer the value derived using the DIB since the Na~D line may have
been saturated. } \newline

{\footnotesize References: [1] \citealt{1997MNRAS.290..303B}; [2]
\citealt{1993A&A...280L...1S}; [3] \citealt{1977ApJ...211..872O}; [4]
\citealt{1983PASP...95..391W}  [5] \citealt{2002ApJ...570..277K}; [6]
\citealt{1997A&A...318..179D}; [7] \citealt{2000ApJ...539L..37H}; [8]
\citealt{1992ApJ...397..664C}; [9] \citealt{1994A&A...285..509G}; [10]
\citealt{1998ApJ...499..375O}; [11] \citealt{1989ApJ...344..320V}; [12]
\citealt{1999A&A...348L...9S}; [13] \citealt{2001ApJ...563..229T}; [14]
\citealt{1998MNRAS.300...64H}; [15] Hynes et al.~(2004); [16]
\citealt{2003ATel..196....1G}; [17] \citealt{1978ApJ...221L..63G}; [18]
\citealt{2001ApJ...555..489O}; [19] \citealt{2002MNRAS.331..169H}; [20]
\citealt{1999IAUC.7291....2D}; [21] \citealt{1990A&A...238..163C}; [22]
\citealt{1989ApJ...337L..81T}; [23] \citealt{1991ApJ...378..293W}; [24]
\citealt{1993MNRAS.265..834C}}\newline

\end{table}

\subsubsection{Systematics in the radius expansion burst method}
\label{nsuncer}
It has been suggested that the burst flux could be anisotropic
(\citealt{2002A&A...382..947K}). However, given the good agreement
between the Globular Cluster distances and the distances to the LMXBs
in these Globular Clusters derived from the photospheric radius
expansion burst properties (\citealt{2003A&A...399..663K}) that seems
unlikely. Furthermore, \citet{2003ApJ...590..999G} showed that
considering the burst peak fluxes of 61 photospheric radius expansion
bursts in the atoll source 4U~1728--34 the degree of anisotropy in the
burst emission in less than 2 per cent.

Two out of the eight neutron star systems studied by
\citet{2003A&A...399..663K} have a photospheric radius expansion peak
burst luminosity that is lower than that of the other six. This lower
peak luminosity is consistent with the Eddington luminosity limit for
hydrogen rather then helium--rich material for a neutron star mass of
1.4 M$_\odot$. So, for some of the neutron star systems in our list we
could have overestimated the distance by a factor $\sim
\sqrt{1.8}$. If we take a 1:4 ratio as was found by
\citealt{2003A&A...399..663K} this would affect 8 systems in our
sample. As was noted by \citet{2003A&A...399..663K} the two sources
whose peak luminosity is consistent with hydrogen accretion have an
orbital period characteristic for normal LMXBs, rather than
ultra--compact X--ray binaries with periods less than $\sim$1 hour
(which necessarily accrete hydrogen poor material
e.g.~\citealt{verbunt1995}). Because the fraction of ultra--compact
X--ray binaries in Globular Clusters might be higher than in the
field, the ratio between the number of sources with a hydrogen--rich,
low Eddington burst luminosity and the number of sources with a
helium--rich, high Eddington burst luminosity could be higher. It is
unlikely, however, that the distances to {\it all} neutron star LMXBs
in our sample have been overestimated. E.g.~in the long period LMXBs
KS~1731--26 and 4U~1636--53 the photospheric radius expansion burst
was consistent with a helium--rich explosion
(\citealt{2000ApJ...542.1016M},
\citealt{1984PASJ...36..839S}, respectively, see also
\citealt{2000ApJ...544..453C}).

\section{Implications and discussion}

Using the data set compiled above, leaving the distances derived using
method $B$ as they are, keeping the discrepancy in the distance
estimate between method $B$ and the radius expansion burst method in
mind, we investigate the Galactic distribution of neutron star and BH
LMXBs and the peak luminosity of BHs.

\subsection{Galactic distribution of LMXBs}

\citet{vawh1995} investigated the Galactic z--distribution of neutron
star LMXBs and concluded that neutron stars should receive an
asymmetric kick at birth from the fact that the rms value of the
z--distribution was $\sim 1$ kpc. We obtain an rms z--values of 1025
pc and 1125 pc for persistent and transient neutron star LMXBs,
respectively (we round rms z--values to the nearest 25 pc). In this we
have excluded XTE~J2123--058 since with a z--value of -12.4 kpc it
would dominate the outcome and \citet{2002MNRAS.329...29C} show that
the systemic velocity is consistent with it being a halo source. It
can be argued on the basis of the large z values that Cyg~X--2 and
EXO~0748--676 are also halo sources although this is much less
clear. E.g.~\citet{2000MNRAS.317..438K} argue that if Cyg~X--2 is at a
distance of $\sim$11.6 kpc it could have originated in the Galactic
plane. However, when we also exclude Cyg~X--2 and EXO~0748--676 we
find an rms z--value for the persistent and transient neutron star
LMXBs of 700 pc and 850 pc, respectively.  \citet{vawh1995} found an
rms--z value of $\sim$500 pc when they excluded Cyg~X--2 and
EXO~0748--676 from their sample. Since we have used 3.8$\times10^{38}$
erg s$^{-1}$ for the Eddington luminosity for the radius expansion
peak luminosity for all neutron stars, the rms z--value we derived for
neutron star LMXBs corresponds to an upper limit. If we use
2.0$\times10^{38}$ erg s$^{-1}$ as the photospheric radius expansion
peak burst luminosity we find an rms z--value of 775 pc and 850 pc for
persistent and transient neutron star LMXBs, respectively (here we
only excluded XTE~J2123--085, if we also exclude Cyg~X--2 and
EXO~0748--676 we find 550 and 650 pc for persistent and transient
neutron star LMXBs, respectively). I.e.~our findings confirm the
result of \citet{vawh1995}.

In a follow--up paper, \citet{1996ApJ...473L..25W} investigated the
differences between the rms values of the neutron star and BH LMXBs
z--distributions. They found that the rms value for the BHs was
substantially lower than that of the neutron star z--distribution
(more than a factor 2). Using the distances of the BHs given in
Table~\ref{disBHC} we now find an rms--value of $\sim$625 pc (we
took a distance of 2.5 kpc for GRO~J0422+32; we excluded the likely
halo object XTE~J1118+480; \citealt{2001ApJ...556...42W} [although note
that some of the evidence leading to the suggestion that XTE~J1118+480
is a halo object was based on the low rms z--value
\citealt{1996ApJ...473L..25W} found for BH LMXBs]), i.e.~close to the upper limit we
find for the neutron star systems. Increasing the distance of
XTE~J1550--564 with a factor 3 does not significantly increase the rms
value of the BH population. The main reason for the difference between
the findings of \citet{1996ApJ...473L..25W} and our findings is that
the distance estimates for most BHs have gone up. The conclusion drawn
by \citet{1996ApJ...473L..25W} that BHs receive a significantly
smaller kick than neutron stars is no longer tenable.

\begin{figure}
  \includegraphics[width=8cm]{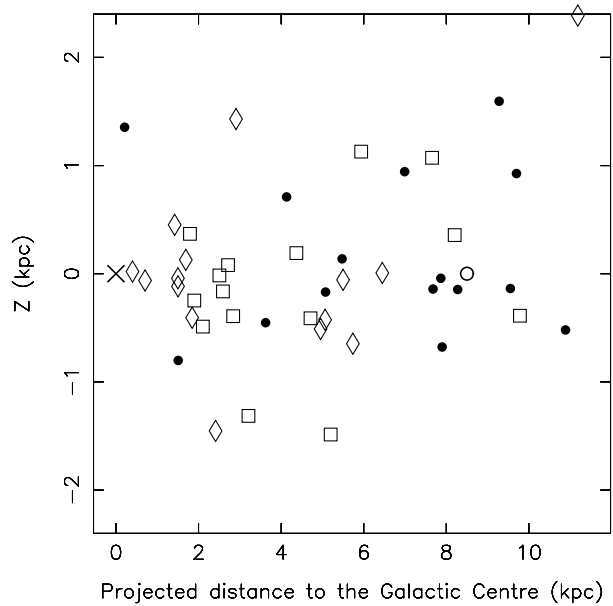}
\caption{The z--distribution of the Galactic persistent (open squares)
and transient (open diamonds) neutron star LMXBs for which a radius
expansion burst has been detected, and BH LMXBs (filled circles) for
which a dynamical mass has been derived as a function of the projected
distance to the Galactic Centre. The projected distance is
$\sqrt{x^2+y^2}$, where $x$ and $y$ are the Cartesian coordinates in
the Galactic plane (see Figure~\ref{xytot}). The location of the Sun
at an assumed distance of 8.5 kpc is indicated with an open circle,
and the location of the Galactic Centre is indicated with a
cross. Four neutron star systems fall off the Figure
(EXO~0748--676, z=-3.1 kpc, XTE~J2123--058, z=-12.4 kpc,
XTE~J1710--281, D$_{GC}=$19.7 kpc, and Cyg~X--2, D$_{GC}=$15 kpc). In
this Figure we use the neutron star distances derived assuming the
Eddington peak burst luminosity was 3.8$\times10^{38}$ erg s$^{-1}$.}
\label{yztot}
\end{figure}

We plotted the z values for transient and persistent neutron star
LMXBs (open diamonds and squares, respectively) and BHs (black dots)
using the distances from Table~\ref{disBHC} and ~\ref{NSdist} as a
function of their projected distance to the Galactic Centre in
Figure~\ref{yztot}. It is interesting that the rms value for the
projected distance to the Galactic Centre for the neutron stars and BH
is 4.8 kpc and 7.0 kpc, respectively (excluding the [likely] halo
sources). Because the Galactic potential in the z--direction decreases
with increasing Galactocentric radius
(e.g.~\citealt{1987AJ.....94..666C}) the larger scale height of BHs can
partly be due to this effect, rather than a kick velocity. E.g.~for
neutron star systems \citet{vawh1995} estimate an rms value of the
z--distribution near the Sun of 650 pc. Furthermore, the symmetric
kick velocity (imparted due to mass loss in the supernova) scales with
the companion mass and the mass lost in the supernova
(e.g.~\citealt{1999A&A...352L..87N}) both can be (much) larger in the
case of BH systems, but it scales inversely with the total mass of the
remaining binary. Lastly, for large asymmetric kick velocities the
binary is more likely disrupted in case of a neutron star than a BH. A
detailed investigation of all the possible explanations for the high
rms value for the BHs is needed in order to draw firm conclusions and
we defer this to a later paper.

\begin{figure*}
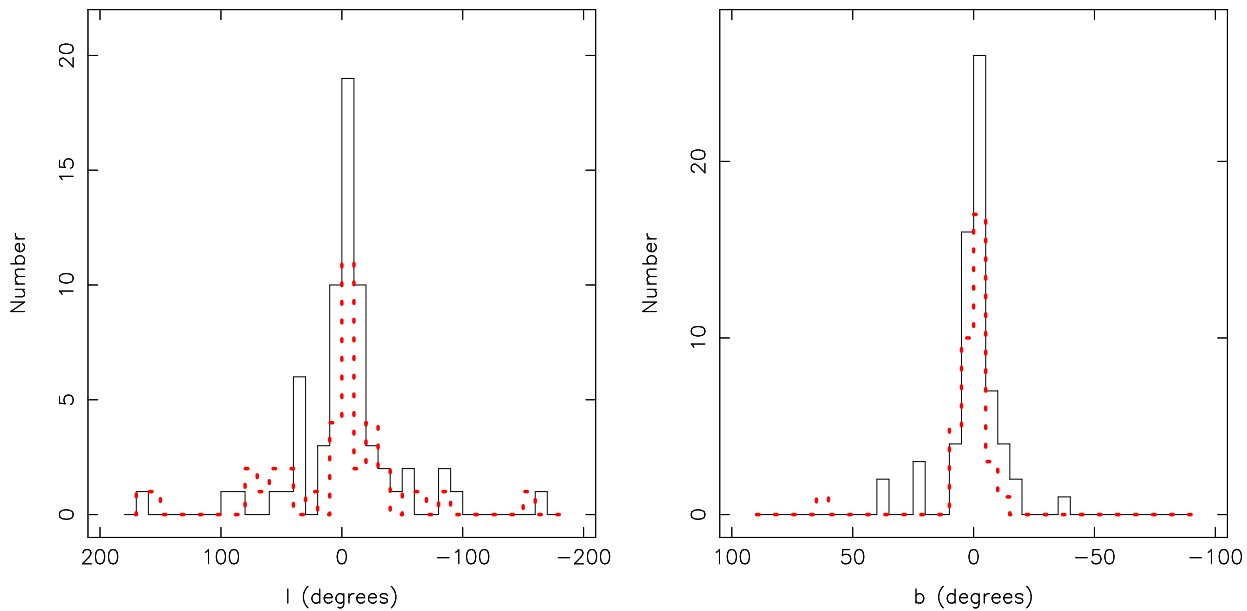

\includegraphics[width=8cm,height=8cm]{l.bhc.ns+6.ps}
\quad
\includegraphics[width=8cm,height=8cm]{b.bhc.ns+6.ps}
\caption{{\it Left panel:} The l  distribution of the neutron
stars (solid line histogram; all sources showing bursts or pulsations
listed in Table~\ref{NSdist} and Liu et al.~2001) in bins 10$^\circ$
wide. The l  distribution of the BHs listed in
Table~\ref{disBHC} and McClintock and Remillard 2004 (dotted line
histogram). {\it Right panel:} The b  distribution for the same
sources and using the same symbols as in the {\it left panel} in bins
5$^\circ$ wide.}
\label{ldist}
\end{figure*}

After \citet{1996ApJ...473L..25W} we also compare the distribution of
the l and b coordinates of the neutron star and BH LMXBs. We first
compare the l and b coordinates of neutron star LMXBs in
Table~\ref{NSdist} with those of the BHs listed in
Table~\ref{disBHC}. A K--S test shows that the probability that the
distributions are the same is 18 per cent for the l coordinate and 52
per cent for the b coordinate (the K--S D--value is 0.33 and 0.24,
respectively. Here and below: the effective number of data point is
always larger than 10 so that the probabilities we quote are quite
accurate, cf.~\citealt{prteve1992}). To investigate this further and to
minimise selection effects, we plot the l and b coordinates of {\it
all} neutron star systems, e.g.~systems where a burst was found, and
pulsars listed in \citet{2001A&A...368.1021L} (solid line histogram in
Figure~\ref{ldist}). We compare this with the BH sources listed in
Table~\ref{disBHC} {\it and} the BH candidates for which no mass
estimate based on a radial velocity study exist (systems in table 3 in
\citealt{2003mcclintock}; the dashed line histogram in
Figure~\ref{ldist}). A one--dimensional K--S test shows that the
probability that the neutron star and BH distributions in l are drawn
from the same distribution is 37 per cent whereas the probability that
b values are drawn from the same distribution is 90 per cent (the K--S
D--value is 0.19 and 0.12, respectively).  Hence, using a larger
sample of neutron stars and BHs makes the probability that the neutron
star and BH distributions are the same larger. The probabilities we
derive are higher than those derived by \citet{1996ApJ...473L..25W}; a
difference we attribute to the fact that \citet{1996ApJ...473L..25W}
had fewer BH systems in their study. We conclude that there is no
evidence for a difference in the l and b distributions of neutron star
and BH LMXBs.

Interestingly, there seems to be an excess of 12 out of a total of 22
BH and an excess of 19 out of a total of 45 neutron star systems with
-30$^\circ<{\rm l}<0^\circ$ over systems with 0$^\circ<{\rm
l}<30^\circ$. There is an excess of 5 out of 35 BH and 13 out of 53
neutron star systems with -10$^\circ<{\rm b}<0^\circ$ and
0$^\circ<{\rm b}<10^\circ$ (see Figure~\ref{ldist}).  The expected
mean number of systems with a positive/negative l of a symmetric
distribution around l=0$^\circ$ is $N/2$, with $N$ the total number of
systems. The variance on this number is $\sqrt{N/4}$.  Comparing this
with the observed number of systems yields significances of the
observed asymmetries of 2.8$\sigma$ for the neutron star distribution
in l, 3.0$\sigma$ for the BH distribution in l, and 3.8$\sigma$ for
the combined neutron star and BH distribution in l. The combined
significance for the neutron star and BH distribution in b is
1.9$\sigma$ only, so we do not discuss this apparent asymmetry in b
further.

The asymmetry in the l--distribution is not an obvious observational
selection effect. For instance, the BeppoSAX WFC monitoring the
Galactic Centre region had a field--of--view of
$40^\circ\times40^\circ$ centred on the Galactic Centre
(\citealt{2001ESASP.459..463I}). The dust maps from \citet{scfida1998}
do not show a large dust asymmetry around the Galactic Centre. It
seems that the neutron star and BH LMXB distribution in l is symmetric
around l=-5$^\circ$. Possibly this asymmetry is related to the bar
structure in the inner part of our Galaxy. It is known that the bar
causes asymmetries in the stellar and gas distributions around (l,b) =
(0,0) (e.g.~see the recent review by
\citealt{2003ASPMerrifield}). Furthermore, \citet{2004MNRAS.349..146G}
found that the distribution of LMXBs follows the stellar mass
distribution, and hence not the spiral arm structure.

\begin{figure*}
  \includegraphics[width=8cm,height=8cm]{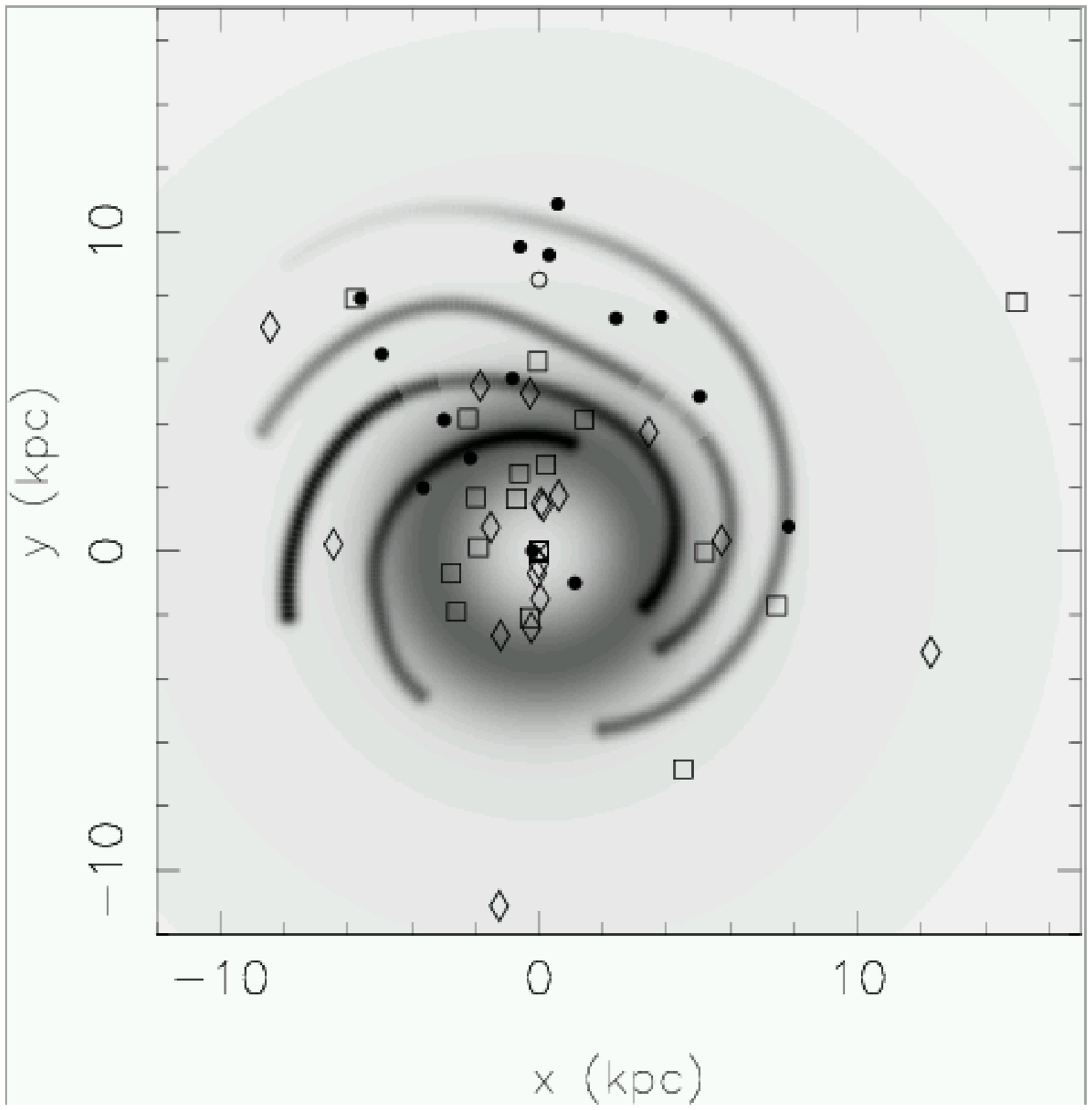}
\quad
\includegraphics[width=8cm,height=8cm]{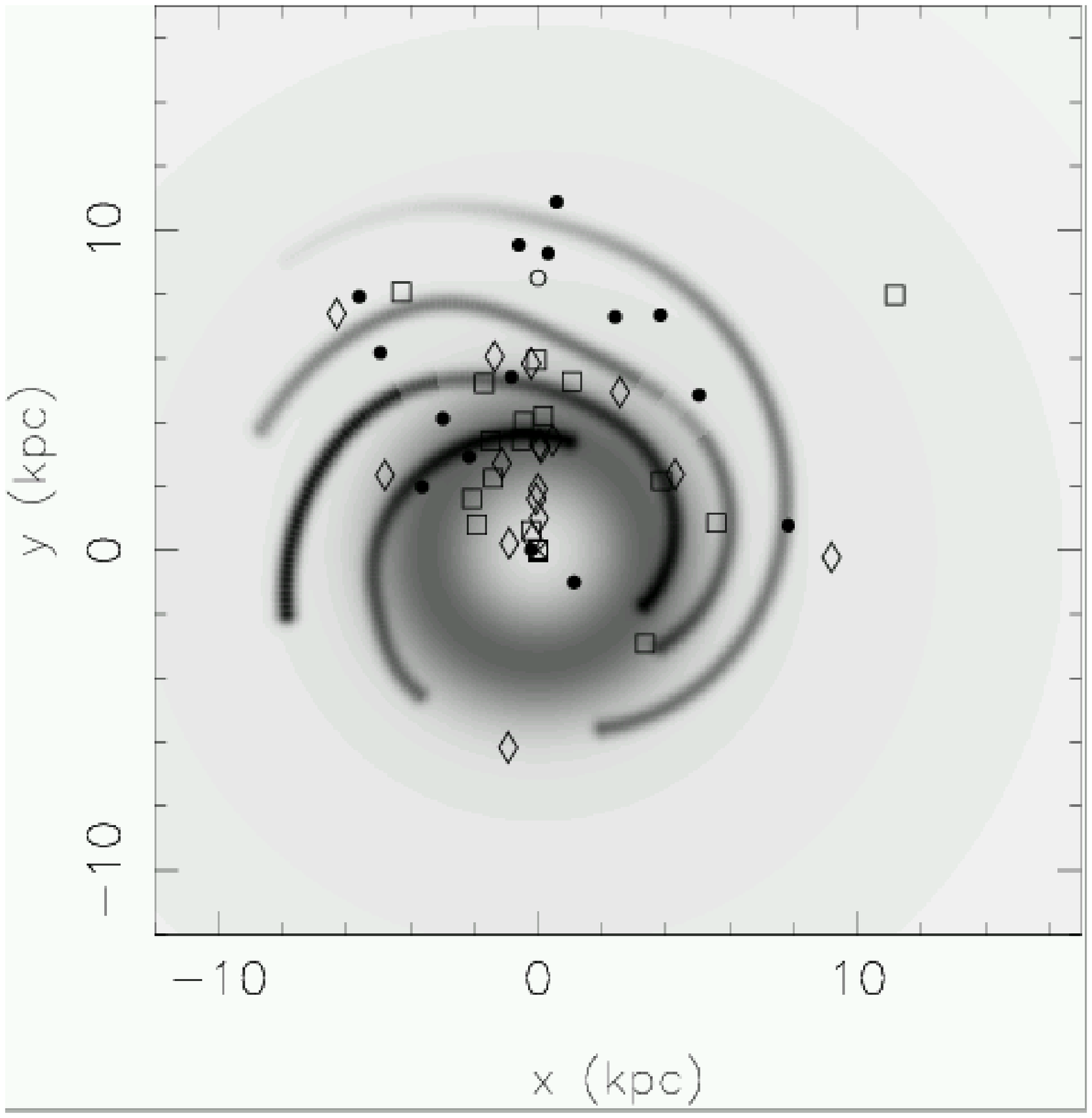}
\caption{{\it Left panel:} The x--y distribution of the Galactic
persistent (open squares) and transient (open diamonds) neutron star
LMXBs for which a photospheric radius expansion burst has been
detected, and BH SXTs (filled circles) for which a dynamical mass has
been derived. The location of the Sun is indicated with an open
circle, and the location of the Galactic Centre at an assumed distance
of 8.5 kpc is indicated with a cross. It is clear that there is a
paucity of nearby neutron star systems compared to nearby BH SXTs. The
neutron star distances were derived assuming that the Eddington peak
burst luminosity was 3.8$\times10^{38}$ erg s$^{-1}$. Overplotted is
the spiral structure and the free--electron density according to the
model of Taylor \& Cordes (1993). {\it Right panel:} same as the {\it
left panel} except that the neutron star distances have been derived
assuming that the Eddington peak burst luminosity was
2.0$\times10^{38}$ erg s$^{-1}$.  }
\label{xytot}
\end{figure*}

Using the distances listed in Table~\ref{disBHC} and~\ref{NSdist} we
also plot in Figure~\ref{xytot} the Galactic x--y distribution of the
neutron stars for which a photospheric radius expansion burst was
observed and the BHs for which a dynamical mass estimate has been
derived. The Galactic spiral arm structure according to the model
described in \citet{1993ApJ...411..674T} has been overplotted. It is
clear that few LMXBs at the other side of the Galactic Centre have
been detected, especially if the Eddington limit for hydrogen rich
material for a 1.4 M$_\odot$ neutron star is applicable to the peak
burst flux of most photospheric radius expansion bursts (see also
\citealt{2002A&A...391..923G}). There also seems to be a paucity of
nearby neutron star LXMBs (this was also noted by
\citealt{2001bhbg.conf..279V}). In order to try to quantify this we
performed a two dimensional Kolmogorov--Smirnov (K--S) test
(\citealt{prteve1992}) to test whether the neutron star and BH
populations are the same. The K--S test gives a D of 0.45 and a
probability that the two populations are the same of 3.8 per cent. If
we decrease the distance of all neutron star systems by a factor
$\sim \sqrt{1.8}$ the probability that the distributions are the same increases
only slightly to 5.2 per cent (D=0.44). It is possible that the
discrepancy in the spatial distribution of the neutron stars and BHs
is a selection effect. Since we only included BHs for which a mass
limit larger than 3$M_\odot$ has been derived, the optical counterpart
must have been detected in quiescence, hence this favours nearby
systems.

\subsection{The black hole outburst peak luminosity}

The BH distances are important for the maximum observed BH
luminosity. We found a somewhat larger distance for V404~Cyg than
often used previously.  \citet{1994MNRAS.271L..10S} found an upper
limit on the distance of V404 Cyg of 3.7 kpc assuming the peak
outburst source luminosity was Eddington limited taking their 90 per
cent confidence upper limit to the mass of the BH of 15 M$_\odot$.
The distance of 4.0$^{+2.0}_{-1.2}$ kpc we find shows that the maximum
observed luminosity exceeds the Eddington luminosity for a 10
M$_\odot$ BH (the limit often quoted to decide whether a source is a
ULX or not).  The distance derived for XTE~J1550--564 may have been
underestimated by as much as a factor 3: its distance could be as
large as 3$\times5.3=15.9$ kpc, although the uncertainties are
large. For this distance the outburst peak luminosity in the 2--20 keV
band alone would be 7$\times10^{39}$ erg cm$^{-2}$ s$^{-1}$ (taking
the flux from
\citealt{somcre1999}). For 4U~1543--47, SAX~J1819.3--2525 and GRS~1915+105, 
systems for which the distance was determined by method $B$, $B$, and
$C$, respectively, it was noticed earlier that the outburst peak
luminosities are super--Eddington for the best--fit BH masses
(cf.~\citealt{2002A&A...391.1013R};
\citealt{2003mcclintock}). \citet{2003ApJ...591..388G} show that the
peak outburst of GRO~J1009--45 most likely also exceeded the Eddington
luminosity for a for $\sim10\,{\rm M_\odot}$ BH (even if the distance
is 5.7 instead of the 9 kpc they favoured). It seems therefore likely
that there are several sources in our own Galaxy which we would
classify as transient ULXs had we observed them in other Galaxies. For
these sources a mass determination has shown that they are not
intermediate mass BHs but rather 5--15 M$_\odot$ BHs. Furthermore, the
fact that LMXB sources (an old population) seem to be capable of
producing super--Eddington luminosities could also help explain the
presence of ULXs in elliptical Galaxies. Assuming that most ULXs are
stellar mass BHs the fact that many ULXs are found a few arc seconds
away from the young stellar clusters could be caused by kick
velocities received at BH formation.

\begin{figure}
\includegraphics[width=7cm,angle=-90]{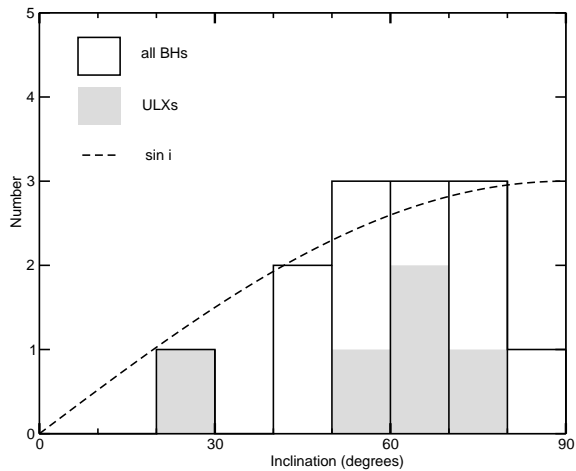}
\caption{Inclination distribution for the dynamically confirmed BH
SXTs (all the inclinations are taken from Orosz 2003). The shaded bins
indicate the five systems which would have been classified as ULX had
we observed them in another Galaxy at their outburst peak
luminosity. XTE~J1550--564 ($i\sim72\pm5^\circ$) is possibly also an
ULX (see text).}
\label{incli}
\end{figure}

If super--Eddington luminosities were to be explained by effects of
``mild--beaming'' (cf.~\citealt{2001ApJ...552L.109K}) then one would
naively expect the inclinations of the sources with the highest
luminosities to be lowest. In Figure~\ref{incli} we plot the
inclination distribution for 13 BH SXTs in Figure~\ref{incli} (the
inclinations for GX~339--4 and XTE~J1859+226 are not well known and
those sources have therefore not been included; inclinations from
\citealt{2002Inventorosz}). Although the amount of sources is low we see
that the inclination distribution of the BH ULX sources is not
skewed to low inclinations but instead clusters around the fiducial
60$^\circ$ point. We did not include XTE~J1550--564
($i\sim72\pm5^\circ$) and GRO~J0422+32 ($i\sim41\pm3^\circ$) as ULXs.

\section{Conclusions}

We studied the distance estimates for BH SXTs in detail. We note that
the uncertainties in the distance estimates due to e.g.~uncertainties
in the spectral type (i.e.~temperature) of the companion star are
probably large. Comparing the distances derived for the neutron stars
Cyg~X--2 and XTE~J2123--058 using the distance estimation method $B$,
which is used for most BH SXTs, with the photospheric radius expansion
burst method, we find that the latter gives larger distances. This
could mean that for some reason method $B$ systematically
underestimates the distance. Possibly this is related to an erroneous
spectral classification of the companion star caused by its fast
rotation. If this is indeed the case this would have important
consequences for the reported difference in quiescent X--ray
luminosities of BH and neutron star SXTs, for the BH SXT peak outburst
luminosities, for the BH masses, and for the Galactic distribution of
BH LMXBs. We find that the distance towards XTE~J1550--564 may have
been underestimated by as much as a factor 3 because the interstellar
extinction could have been overestimated in that case. As was noticed
before, several BH SXTs observed in our Galaxy would be classified as
ULXs had we observed them in another Galaxy (at least 5 but perhaps
even 7 out of the 15 dynamically confirmed BH SXTs). This suggests
that many (transient) ULXs in other Galaxies could well be stellar
mass BHs.  A re--evaluation of the distance to the Galactic plane of
neutron star and BH LMXBs shows that there is no longer evidence for a
smaller rms--value of the z--distribution for BH systems. Such a
difference had been interpreted as evidence for the absence of
asymmetric kicks during BH formation. However, before firm conclusions
can be drawn about the similarities or differences between neutron
stars and BH kicks the details of the formation and Galactic
distribution have to be investigated. Finally, we found that the l
distribution of Galactic LMXBs is asymmetric around l=0$^\circ$.

\section*{Acknowledgments}  \noindent  Support for this work was provided by
NASA through Chandra Postdoctoral Fellowship grant number PF3--40027 awarded by
the Chandra X--ray Center, which is operated by the Smithsonian Astrophysical
Observatory for NASA under contract NAS8--39073. GN is supported by PPARC. The
authors would like to thank the referee for his/her valuable comments which
helped improve the manuscript considerably. We would like to thank Hans--Jakob
Grimm, Jeff McClintock, Jim Pringle, and Frank Verbunt for useful discussions
and Duncan Galloway for sharing results before publication. This research made
use of results provided by the ASM/RXTE teams at MIT and at the RXTE SOF and
GOF at NASA's GSFC. The research has made extensive use of NASA's Astrophysics
Data System. 

\bibliographystyle{mn_new}

\end{document}